\documentclass{book}    % fundamental class file is 'book'.
\usepackage{piers}  % use the file 'piers.sty'.
\begin{document}

%=== Input Title here ================================
\title{Casimir Friction for  Media of Finite Density}
\maketitle
%===================================================

%=== List of authors (in order) ========
%-- Author(s) for the first affiliation ---
\author      {Johan S. H{\o}ye}
\affiliation {Department of Physics, Norwegian University of Science and Technology}
\address     {}% optional
\city        {Trondheim}
\postalcode  {}% optional
\country     {Norway}
\phone       {345566}    % optional
\fax         {233445}    % optional
\email       {johan.hoye@ntnu.no}  % optional
\misc        { }  % optional
\nomakeauthor
%------------------------------------

%=== List of authors (in order) ========
%-- Author(s) for the second affiliation ---
\author      {Iver Brevik}
\affiliation {Department of Energy and Process Engineering, Norwegian University of Science and Technology}
\address     {}% optional
\city        {Trondheim}
\postalcode  {}% optional
\country     {Norway}
\phone       {345566}    % optional
\fax         {233445}    % optional
\email       {iver.h.brevik@ntnu.no}  % optional
\misc        { }  % optional
\nomakeauthor
%-------------------------------------
%=== List of other authors (in order) ========
%......

%---Output of Authors----------------------
\begin{authors}

{\bf Johan S. H{\o}ye}$^{1}$ and  {\bf Iver Brevik}$^{2}$, \\
\medskip
$^{1}$Department of Physics, Norwegian University of Science and Technology, Trondheim, Norway\\
$^{2}$Department of Energy and Process Engineering, Norwegian University of Science and Technology, Trondheim, Norway

\end{authors}
%--------------------------

%---Content of Paper Abstract-----------------------
\begin{paper}

\begin{piersabstract}
This work is a continuation of our  papers from the last couple of  years on the Casimir friction for a pair of particles at low relative velocity. The new element in the present analysis  is to allow  the media to be  dense. Then the situation becomes more complex due to induced dipolar correlations, both within the two planes, and between the planes. We show that the structure of the problem can be simplified by regarding the  two plates to be a generalized version of a pair of particles.  The force is predicted to be very small, far beyond  what is practically measurable.

\end{piersabstract}

%---Content of Paper Text-----------------------
\psection{Introduction}

The typical situation envisaged in connection with Casimir friction is the one where  two parallel semi-infinite dielectric nonmagnetic plates at micron or semi-micron separation are moving longitudinally  with respect to each other, one plate being at rest, the other having a nonrelativistic velocity $\bf v$. Usually the plates are taken to have the same composition, their permittivity  $\varepsilon(\omega)$  being frequency dependent.

Most previous works on Casimir friction are formulated within the framework of macroscopic electrodynamics. Some references in this direction are \cite{pendry97,pendry10,volokitin07,volokitin11,dedkov11,dedkov12,philbin09}. In particular, the application of the theory to graphene materials is a very promising avenue of approach; cf., for instance, Ref.~\cite{volokitin11}. In the present paper we focus on the following themes:

$\bullet$ We make use of {\it statistical mechanical methods} for harmonic oscillators,  moving
with respect to each other with constant velocity $\bf v$, at a finite temperature $T$.  We claim that such a strategy is  quite powerful. We have used this method repeatedly in previous recent  investigations \cite{hoye10,hoye11A,hoye11B,hoye12A,hoye12B}; cf. also the earlier  papers \cite{hoye92,brevik88} in which the foundations of the method were spelled out. The essence of the method is to generalize the statistical mechanical Kubo formalism to time-dependent cases.

$\bullet$ These methods are then used to generalize the theory to the case of {\it dense} media. This is a nontrivial task,  as the additivity property   holding for dilute media is no longer valid.  One will have to deal with a more complicated form of the Green function. The atomic polarizabilities appearing in the theory of dilute media have to be replaced by by functions based upon the frequency dependent permittivity.

$\bullet$ It turns out that  the friction force becomes finite, in a mathematical sense, although extremely small. Unless an enormous enhancement can be devised, the Casimir friction, where the media are not in direct contact, appears to be a purely academic effect.

 We mention that the microscopic approach has been followed by other investigators also, especially by Barton \cite{barton10A,barton10B,barton11}. The equivalence between our approach and that of Barton is actually not so straightforward to verify, but has been shown explicitly \cite{hoye11B}.

\psection{Dilute media}

 For a pair of  polarizable particles the electrostatic dipole-dipole pair interaction perturbs the Hamiltonian by an amount
\begin{equation}
-AF(t)=\psi_{ij}s_{1i}s_{2j}, \label{1}
\end{equation}
where the summation convention for repeated indices $i$ and $j$ is implied. The $s_{1i}$ and $s_{2j}$ are  components of the fluctuation dipole moments of the two particles $(i,j=1,2,3)$. With  electrostatic dipole-dipole interaction we can write
\begin{equation}
\psi_{ij}=-\frac{\partial^2}{\partial  x_i\partial x_j}\psi, \quad \psi=\frac{1}{r}, \label{2}
\end{equation}
(i.e. $\psi_{ij}=-(3x_ix_j/r^5-\delta_{ij}/r^3)). $ Here ${\bf r}={\bf r}(t)$ with components $x_i=x_i(t)$ is the separation between the particles. The time dependence in Eq.~(\ref{1}) is due to the variation of $\bf r$ with time $t$, and the interaction will vary as

\begin{equation}
-AF(t)=\left[ \psi_{ij}({\bf r}_0)+\left(\frac{\partial}{\partial x_l}\psi_{ij}({\bf r}_0)\right)v_l t+...\right]s_{1i}s_{2j}, \label{3}
\end{equation}
where $v_l$ are the components of the relative velocity $\bf v$. The components of the force $\bf B$ between the oscillators are
\begin{equation}
B_l= -T_{lij}s_{1i}s_{2j}, \quad T_{lij}=   \frac{\partial}{\partial x_l}\psi_{ij}. \label{4}
\end{equation}
The friction force is due to the second term of the right hand side of Eq.~(\ref{3}), and for dilute media the first term can be neglected.

For the time dependent part of Eq.~(\ref{3}) we may write $-AF(t)\rightarrow -A_lF_l(t)$ where $A_l=B_l$ and $F_l(t)=v_lt.$ According to the Kubo formula the perturbing term leads to a response in the thermal average of $B_l$ given by
\begin{equation}
\Delta \langle B_l(t)\rangle=\int_{-\infty}^\infty \phi_{BAlq}(t-t')F_q(t')dt', \label{5}
\end{equation}
where the response function is ($t>0 $)
\begin{equation}
\phi_{BAlq}(t)=\frac{1}{i\hbar}{\rm{Tr}}\,\{ \rho[A_q, B_l(t)]\}. \label{6}
\end{equation}
Here $\rho$ is the density matrix and $B_l(t)$ is the Heisenberg operator $B_l(t)=e^{itH/\hbar}B_l\,e^{-itH/\hbar}$ where $B_l$ like $A_q$ are time independent operators. With Eqs. (\ref{3}) and (\ref{4}) expression (\ref{6})  can be rewritten as
\begin{equation}
\phi_{BAlq}(t)= T_{lij}T_{qnm} \phi(t)\delta_{in}\delta_{jm},     \label{7}
\end{equation}
\begin{equation}
\phi(t)\delta_{in}\delta_{jm}={\rm{Tr}}\,\{\rho\frac{1}{i\hbar}[s_{1i} s_{2j}, s_{1n}(t) s_{2m}(t)]\}
\label{7a}
\end{equation}
(i.e. the situation with scalar polarizability is assumed such that $\langle s_{ai} s_{an}(t)\rangle=0$ for $i\neq n$). Further following Refs.~\cite{hoye12A} and \cite{hoye13} one can introduce the correlation function $g(\lambda)$ in imaginary time $\lambda=it/\hbar$ where
\begin{equation}
\phi(t)=\frac{1}{i\hbar}[g(\beta+\lambda)-g(\lambda)]
\label{8}
\end{equation}
\begin{equation}
g(\lambda)\delta_{in}\delta_{jm}={\rm Tr}[\rho s_{1n}(t)s_{2m}(t)s_{1i}s_{2j}] \label{9}
\end{equation}
with Fourier transforms
\begin{equation}
\tilde\phi(\omega)=\int_0^\infty \phi(t)e^{-i\omega t}\,dt \quad\mbox{and}\quad \tilde g(K)=\int_0^\beta g(\lambda) e^{iK\lambda}\,d\lambda.
\label{9a}
\end{equation}
Here $K=i\hbar\omega$ and  $\beta=1/(k_B T)$, where $k_B$ is Boltzmann's constant and $T$ the temperature. Then one has \cite{brevik88}
\begin{equation}
\tilde\phi(\omega)=\tilde g(K).
\label{10}
\end{equation}
Further we now have
\begin{equation}
g(\lambda)=g_1(\lambda)g_2(\lambda)\quad\mbox{and}\quad  \tilde g(K)=\frac{1}{\beta}\sum_{K_0}\tilde g_1(K_0)\tilde g_2(K-K_0)
\label{11}
\end{equation}
where for a simple harmonic oscillator with zero frequency polarizability $\alpha_a$ and eigenfrequency $\omega_a$ ($a=1,2$)
 \begin{equation}
{\tilde g}_a(K)=\alpha_{aK}=\frac{\alpha_a(\hbar \omega_a)^2}{K^2+(\hbar \omega_a)^2}. \label{15}
\end{equation}
Altogether following Ref.~\cite{hoye12A} the friction force is then given by
\begin{equation}
F_{fl} =-iG_{lq}v_q\frac{\partial\tilde\phi(\omega)}{\partial\omega}\Big|_{\omega=0}= -G_{lq}v_q H\frac{\pi \beta}{2}\delta(\omega_1-\omega_2), \label{17}
 \end{equation}
 where
 \begin{equation}
 G_{lq}=T_{lij}T_ {qij}, \quad     H=\left( \frac{\hbar \omega}{2\sinh (\frac{1}{2}\beta \hbar \omega)}\right)^2 \alpha_1\alpha_2, \label{18}
 \end{equation}
 and $\omega_1-\omega_2=\omega$.

The treatment above can be extended to more general polarizability that can be written as ($a=1,2$)
\begin{equation}
\alpha_a(K)=\int\frac{\alpha_{Ia}(m^2)m^2}{K^2+m^2}\,d(m^2).
\label{12}
\end{equation}
This will generalize Eq.~(\ref{17}) to oscillators with a band of eigenfrequencies.

Finally by integrating $G_{lq}$ over space one obtains for dilute media the friction force  $F$ (per unit area) between two half-planes that move parallel to each other \cite{hoye12A}
 \begin{equation}
  F=-GvH,   \quad        G=\frac{3\pi}{8d^4}\rho_1\rho_2,        \label{19}
 \end{equation}
where now with $m=\hbar\omega$
\begin{equation}
H=\frac{\pi\beta\hbar}{2}\int\frac{m^4\alpha_{I1}(m^2)\alpha_{I2}(m^2)}{\sinh^2{(\frac{1}{2}\beta m)}}\,dm.
\label{13}
\end{equation}
Here $v$ is the relative velocity in the $x$ direction. Moreover
 $\rho_1$ and $\rho_2$ are the particle densities in the half-planes, and $d$ is the separation between the half-planes.

 When  $v$ is low, there is according to (\ref{18}) no friction. Physically this is  understood by the circumstance that excitations of the quantized system require disturbances with frequencies matching the energy difference $\hbar(\omega_1-\omega_2)$. Low constant velocity represents the limit of {\it zero} frequency.

\psection{Dense media}

For higher densities, separate oscillators both within each plane and between planes will be correlated. This will add to the complexity of the problem. Some simplification  can be achieved, however,   by regarding the two half-planes as a generalized version of a pair of particles. A detailed exposition of the theory for this case is given in \cite{hoye13}. Here, we sketch some points.

Expression (\ref{9}) is a thermal average of four oscillating dipole moments. They have Gaussian distributions since they represent coupled harmonic oscillators. This means that averages can be divided into averages of pairs of dipole moments. To better see the structure of these correlations one-dimensional oscillators with interaction energy $\phi s_1 s_2$ were considered \cite{hoye12A}. Then non-zero averages $\langle s_1 s_2\rangle$ as well as $\langle s_1^2\rangle$ and $\langle s_2^2 \rangle$ could be evaluated from which
$\langle s_1 s_2s_1 s_2\rangle-\langle s_1 s_2\rangle\langle s_1 s_2\rangle=\langle s_1^2\rangle\langle s_2^2 \rangle+\langle s_1 s_2\rangle\langle s_1 s_2\rangle$ and its structure was obtained. This was further extended to obtain the expression and the structure for $\langle s_1(t) s_2(t) s_1 s_2\rangle-\langle s_1(t) s_2(t)\rangle\langle s_1 s_2\rangle$. Then the $\phi$ was replaced by the electrostatic interaction (\ref{1}) and (\ref{2}), and the two particles were replaced by the two half-planes. The corresponding correlation function is then found by the Green function solution of Maxwells equations for the electrostatic problem \cite{hoye98}. At the end of this evaluation the effect of mutual correlations ($\langle s_1 s_2\rangle\neq 0$) could be neglected, keeping only pair correlations within each half-plane. The result of all this was then that the low density expressions of the previous section are kept except that the polarizability is replaced by
\begin{equation}
4\pi \rho_a\alpha_{aK} \rightarrow \frac{2(\varepsilon_a-1)}{\varepsilon_a+1}.      \label{26}
\end{equation}
where $\varepsilon_a$ is the dielectric constant or relative permittivity where for low density $\varepsilon_a-1=4\pi\rho_a\alpha_a$.

Assume now that the plates are equal, and assume the Drude model for the permittivity ($\varepsilon_a=\varepsilon$)
\begin{equation}
\varepsilon=1+\frac{\omega_p^2}{\zeta(\zeta+\nu)}, \label{28}
\end{equation}
where $\zeta=i\omega$, and where $\nu$ represents damping of plasma oscillations due to finite conductivity of the medium. With this one finds
\begin{equation}
\frac{\varepsilon-1}{\varepsilon+1}=\frac{q^2}{K^2+q^2+\sigma|K|}
\label{30}
\end{equation}
where $q^2=(\hbar\omega_p)^2/2$ and $\sigma=\hbar\nu$. The physical interpretation of this is that the electron plasma acts as set of damped harmonic oscillators that all have the same eigenfrequency $\omega_p/\sqrt{2}$ (replacing the zero eigenfrequency of expression (\ref{28}). This is the eigenfrequency of surface plasma waves.

With relations (\ref{28}) and (\ref{30}) we can repeat the calculations that led to the friction force per unir area $F$ (\ref{19}). Then the frequency distribution $\alpha_I(m^2)$ of Eq.~(\ref{13}) is needed. With expression (\ref{30} it is (for small $\nu$) \cite{hoye12A}
\begin{equation}
m^2\alpha(m^2)=\frac{q^2}{2\pi\rho}\frac{\sigma q}{x^2+(\sigma q)^2}, \quad x^2=m^2-q^2.
\label{31}
\end{equation}
With this some calculation leads to the follwing force expression per unit area ($\rho_1=\rho_2=\rho$)
\begin{equation}
F=-\frac{3kTv}{128\pi \nu d^4}\left( \frac{\frac{1}{2}\beta q}{\sinh (\frac{1}{2}\beta q)}\right)^2. \label{29}
\end{equation}
It is noteworthy that the particle density $\rho$ does not occur explicitly in this expression.
It is present  indirectly, though, in the plasma frequency $\omega_p$ which decreases with decreasing density with the consequence that $F$ even increases with decreasing density due to the $\sinh(\cdot)$ term.

Also it may not be so obvious why $F$ is inversely proportional to $\nu$ that represents damping of plasma oscillations due to finite conductivity. The point here is that $\nu$ represents the width of the frequency spectrum of plasma oscillations. For small $\nu$ the friction is due to overlapping frequencies, and small $\nu$ means narrow frequency band and large overlap of frequencies.

Consider, as an example, gold at $T=300$ K, corresponding to $kT=25.86$ meV. Then $\hbar \omega_p=9.0$ eV, $\hbar \nu=35$ meV,  $ q=\hbar \omega_p/\sqrt{2}=6.36$ eV, $\frac{1}{2}\beta q=123$. We can then calculate  the first factor in the expression (\ref{29}).  Choosing $v=100$ m/s for the relative velocity and a small separation $d=10$ nm between the plates, the first factor in (\ref{29}) becomes 5.81 mPa. The second factor in (\ref{29}), however, containing the $\rm sinh(\cdot)$ term, washes the friction force out for all practical purposes.

\psection{Conclusion}

There have been various approaches to the Casimir friction problem in the literature, and they are actually  quite difficult to compare. The statistical mechanical approach which we have presented above, can be related to that  of Barton since they both use a microscopic description \cite{barton10A,barton10B,barton11}. As mentioned above, we have shown in an earlier work that our approaches lead to the same result \cite{hoye11B}.

In the case of dilute media, the friction force is given by the expression (\ref{17}). For low velocities $v$, the force is zero. For high velocities, making the disturbance large enough to excite frequencies comparable to the excitation frequencies for the molecules, the friction force is finite.

For dense media, the force expression (\ref{29}) is overwhelmingly suppressed by the $\rm sinh(\cdot)$ factor. It is far beyond measurability for typical metals.

In Ref.~\cite{hoye13}, we made an attempt to compare our results with related ones of  Volokitin and Persson \cite{volokitin07}, and the interested reader may consult that source.  Their approach was within macroscopic electrodynamics, a method quite different from that followed by us above.

Note added in the proof: A remark should be added concerning the numerics: Recent considerations have shown that it is the case of low frequencies that is of primary importance here. Taking that into account, the expression (\ref{29}) will be replaced by a different, and much larger, estimate. Our analytical considerations remain however unchanged.

%\ack

\end{paper}
%--------------------------

\end{document}